\documentstyle[11pt]{article}
\begin{document}
\title{Correlation effects on transport through few-electrons systems}
\author{J. J. Palacios$(^\ast )$, L. Martin-Moreno$(^{\ast \ast })$ and
C. Tejedor$(^\ast )$ \\
$(^\ast )$ Departamento de F\'{\i}sica de la Materia Condensada. \\
Universidad Aut\'onoma de Madrid.  \\
Cantoblanco, 28049, Madrid. Spain. \\
$(^{\ast \ast })$ Instituto de Ciencia de Materiales (CSIC) \\
Universidad Aut\'onoma de Madrid. \\
Cantoblanco, 28049, Madrid. Spain.}
\maketitle

\begin{abstract}
We study lateral tunneling through a quantum box
including electron-electron interactions in the presence of a
magnetic field which breaks single particle degeneracies. The
conductance at zero temperature as a function of the Fermi energy in the
leads consists of a set of peaks related to changing by one
the electron occupancy in the box. We find that the
position and heights of the peaks are
controlled by many-body effects. We compute the conductance up to 8
electrons for several cases where correlation effects
dominate. In the range of intermediate fields spin selection rules
quench some peaks. At low and high fields the behavior of the conductance
as a function of the number of electrons is very different due to
big changes in the many-body ground state wavefunctions.
\end{abstract}

\newpage

\baselineskip 20pt

The increasing ability for producing extremely small cavities where only a
few electrons ($N \leq 10$) coexist, allows the study of many-body effects
on the lateral tunneling through such structures\cite{ptod}.
The concept of capacitance is too simple to describe effects,
like exchange and correlation, which should be extremely important in
those small systems. In many cases, the experiments are performed in the
presence of a high magnetic field $\vec {B}$.The aim of this paper is
to show that correlation effects are crucial, even in the case in
which the system is not in the fractional quantum Hall regime.
A Keldysh framework \cite{513} is used to obtain the conductance
of a square quantum box, in the linear response regime, including
electron-electron interaction. The leads are assumed to behave like a
Fermi liquid, {\em i.e.}, electrons there can be described in terms of
quasiparticles. Hartree, exchange and correlation interactions among
all the electrons confined in the box are included in order to calculate
conductances. We include a cut-off in the two-dimensional Coulomb interaction
to take into account the finite width of actual systems.

We analyze lateral magnetotunneling including many body effects
in the dot by using a Keldysh formalism \cite{513},\cite{514},\cite{512}.
We take two equal barriers separating the box from each lead.
So, the current is given by \cite{513}
\begin{eqnarray}
J= \frac{-2e}{h} \int d \omega \left[ f _ L ( \omega )-f _ R (
\omega ) \right]
Im \left[ tr \{\Gamma\;G^r \} \right] \nonumber
\end{eqnarray}
where
$f _L$ and $f _R $ are the Fermi distributions of the left and right
leads respectively and $G^r$ is the non-equilibrium retarded
Green's function including all the many body effects as well as the
coupling to the leads given by $\Gamma$.
In a single particle basis,
\begin{eqnarray}
 G ^r _ {i,j} (t)=-i \theta (t) \left( \langle d _ i
(t) d _ j ^{\dagger} \rangle + \langle d _ j ^{\dagger} d _i (t)
\rangle \right) \nonumber
\end{eqnarray}
and
\begin{eqnarray}
\Gamma _ {i,j} ( \omega )=2 \pi \sum _ {l} \rho _ l ( \omega )
\langle i | V | l \rangle \langle l | V | j \rangle. \nonumber
\end{eqnarray}
$d _i$ and $d _j ^{\dagger} $ are the annihilation and creation
operators of box states $|i \rangle $ and $|j \rangle $
respectively, the angular brackets in $G ^r _ {i,j} $ mean thermal
average, $V$ is the potential that couples the box to the leads,
these having eigenstates $|l \rangle $ and density of states $\rho
_ l$. It is important to realize that the coupling $\Gamma$ can be
obtained as output of a single-particle calculation of tunneling.
We concentrate in the linear regime at zero temperature so that
thermal averages become expectation values in the ground state,
$G^r$ being the equilibrium retarded Green's function at
zero temperature. The total Hamiltonian is represented in the basis of
antisymmetrized configurations $\alpha ^{(N)}
\equiv \{ n _i \} $ where $n _ i$ are the occupations of single
particle states $|i \rangle $ verifying $n _ i =0 $ or $1$
and $\sum _ {i} n _ i =N$. The many-body eigenstates with energies $E ^{(N)}
_ \beta $ of a box with $N$ electrons
are written as $ \Phi ^{(N)} _ \beta = \sum _ {\alpha } c _
{\alpha , \beta } \alpha ^{(N)} $. In this basis, the equilibrium
Green's function  $g^r$  for an isolated box takes,
in the Lehmann representation, the form
\begin{eqnarray}
g ^{r (N)} _{i,j} (\omega )= \lim _ {\eta \rightarrow 0 } \sum _ {\beta
} \left[
\frac{ \Delta _ {i,j} ^{(N) \beta +}} { \omega + E ^{(N)} _ 0 - E ^{(N + 1)}
_ \beta +i \eta } +
\frac{ \Delta _ {i,j} ^{(N) \beta -}} { \omega - E ^{(N)} _ 0 + E ^{(N - 1)}
_ \beta +i \eta } \right] \nonumber
\end{eqnarray}
where the numerators are spectral weights
$\Delta _ {i,j} ^{(N) \beta +} =
\langle \Phi _ 0 ^{(N)} | d _i | \Phi _ \beta ^{(N+1)} \rangle
\langle \Phi _ \beta ^{(N+1)} | d _j ^+ | \Phi _ 0 ^{(N)} \rangle $
and $\Delta _ {i,j} ^{(N) \beta -} =
\langle \Phi _ 0 ^{(N)} | d _i ^+ | \Phi _ \beta ^{(N-1)} \rangle
\langle \Phi _ \beta ^{(N-1)} | d _j | \Phi _ 0 ^{(N)} \rangle $.
The calculation of the conductance $G=eJ/ \Delta \mu $, requires
the coupling to the leads by using a selfenergy $\Sigma^r =
(g^r)^{-1}-(G^r)^{-1}$.
The interaction and the coupling to the leads must be solved simultaneously.
Such an analysis has been only done for the Anderson
Hamiltonian\cite{512},\cite{5prl611768},\cite{5prl663048}.
A Kondo-like peak appears in the density of states at the Fermi energy and at
zero
temperature due to correlations to the leads. This would
give rise to a perfect transparency.
The existence of the Kondo effect requires degeneracy, usually of spin, of
the single-particle levels. If the degeneracy is broken by
the Zeeman term due to magnetic field, the Kondo peak
shifts away from the Fermi energy and, again, correlations to the leads
are not important for the properties at such energy\cite{5prl611768}.
Therefore, in our problem with magnetic field, we neglect correlations
in the coupling to the leads and consider the selfenergy
$\Sigma^r$ only as the non-interacting, single-particle
selfenergy $\Sigma^{sp}$.
A very good approximation for the single-particle selfenergy is to
consider that the coupling to the leads only broadens the levels $i,j$
but does not shift or mix them, {\em i.e.}
$\Sigma ^{sp} _ {i,j} \simeq -i \Gamma _ {i,j} \delta _ {i,j} $.
Then, the conductance in the presence of the magnetic field becomes
\begin{eqnarray}
G= \frac{e ^2}{h} \sum _ {N,i} \left[ \frac{ \Delta _ {i,i} ^{(N)+}
\Gamma _ {i,i} ^2 ( E _F )} { \left( E _F - \delta V- E ^{(N + 1)} _ 0
+ E ^{(N)} _ 0 \right) ^2+  \Gamma _ {i,i} ^2 }
+ \frac{ \Delta _ {i,i} ^{(N)-}
\Gamma _ {i,i} ^2 ( E _F )} { \left( E _F - \delta V+ E ^{(N - 1)} _ 0
- E ^{(N)} _ 0 \right) ^2+  \Gamma _ {i,i} ^2 } \right] \nonumber
\end{eqnarray}
where, for zero temperature and well resolved resonances, the
only significant contribution comes from $ \beta \equiv 0$.
$E_F$ is the Fermi level of the leads and $\delta V$ gives the bottom
of the box potential with respect to those of the leads.
The conductance reduces to a set of peaks, each one related to
the variation of the discrete number $N$. This is achieved when the
Fermi level at the leads verifies $E _ F \approx \delta V \pm E ^{(N \pm
1)} _ 0 \mp E ^{(N)} _ 0 $. The position of each peak of the
conductance is a many-body feature, a result also
obtained\cite{514},\cite{515} in the very different regime
$k_B T \gg \Gamma $. Correlation effects reflect
on the height of each peak\cite{514}, which is given by $\sum _ {i}
\Delta _ {i,i}^{(N)}$. The width of each peak is given by the single particle
coupling $\Gamma _ {i,i}$.

We apply the above discussed scheme to a square box defined
by two barriers built up in a wire (along the $y$ direction)
of width $W$ by means of transversal (i.e. in the $x$ direction)
gate potentials. These gate potentials
create two effective barriers of width $l _ b $ and height
$V _ b$ separated from each other by the distance $W$.
In such a square geometry, only solutions for isolated
boxes with $N=2$ and $B=0$ have been calculated\cite{520}.
We work with a strong
perpendicular magnetic field described in a Landau gauge ($ \vec{A}=Bx
\vec{u _ y} $). Each single particle state will
be labelled as $i \equiv (n, k, \sigma)$ where $\sigma $ is the spin
index and $n$ and $k$ are the discrete
quantum numbers related to spatial shape of
the wavefunction.
These single particle states are straightforwardly obtained from the
diagolization of the Hamiltonian represented in a basis of sines
and cosines in the $x$ and $y$ directions. From these wavefunctions,
the broadenings $\Gamma _{i,i}$, required for the calculation of the
conductance, are obtained using their expressions given before.

We present here results of the calculation for a square box of side
$W=100 nm$. We use its single particle eigenfunctions to describe
electron-electron interactions within the box.
The calculation of each matrix element of the total Hamiltonian
reduces to compute four-dimensional integrals involving single particle
wave functions. By diagonalizing the Hamiltonian one obtains
the energies of $N$ electrons in the box as well as the eigenstates in
the form of linear combinations of configurations.
The box is defined by two barriers of width $l _ b =25
nm$. In order to have weak coupling between the box and the leads
we take a barrier height $V _ b =12meV$.
Following the experimental procedure, we fix the Fermi energy of the
leads at $E_F=11.5meV$ and move the box bottom potential $\delta V$.

There are three regimes of $B$ depending on the characteristics of the
many-body ground states.
First, we present results in the intermediate regime of magnetic
fields. The main result is that the ground state of the box with $N$
electrons change its multiplet symmetry when varying the magnetic field.
This has an important implication:  for some ranges of $B$ the difference
between the total spin of $N$ and $N \pm 1$ electrons is larger than $1/2$.
Then, the spectral weights are zero and the conductance of the box when passing
from $N$ to $N+1$ electrons disappears. Later on, following with the variation
of
the field, the multiplet character of the ground state changes once again
and the difference $1/2$ is restored and the peak of the conductance appears
once again. This is observed in figure 1 in the conductance of our square box
with magnetic fields corresponding to have $5$, $6$ and $7$ flux quanta
through the box (this implies to cover a range between $2$ and $3T$).
Since we fix the energy difference between the top of the barrier and
the Fermi energy of the leads, the broadening of all the peaks is practically
the same. The numbers within the figure stand for the
number of electrons within the box. For the lower field, the conductance
decreases when $N$ increases, due to the decrease of the spectral weights.
When the magnetic field increases, the peak corresponding to $4$ electrons
disappears because the ground state of $3$ electrons has a total spin
$S=3/2$ while the ground state of $4$ electrons has a total spin $S=0$
so that the spectral weight is zero. This gives a zero in the conductance
intensity although a small lowering of the bottom of the dot potential
would allow the introduction of the electron in the dot without implying
current ({\em i.e.} conductance) in the whole system.
For an even higher field, the ground state of the dot with $3$ electrons
recovers $S=1/2$ and the peak conductance corresponding to $4$ electrons
appears once again.
Apart from selection rules as the above discussed, the peak
heights are given by the spectral weights $\Delta _ {i,i}^{(N)}$ obtained
from the $N$-electrons wave functions of the box.
At the higher field, the peak corresponding to $2$ electrons
becomes rather small because the ground state corresponds to single particle
wavefunctions in the outer region of the box
due to Coulomb repulsion while the $1$ electron  ground
state is situated in the inner region so that the spectral function reduces
significantly. This is similar to the orbital angular momentum rule obtained in
circular dots\cite{514},\cite{516}.
The peaks have been shifted with respect to the single particle result
by the charging effect  that, in this case, results to be rather constant
with the number of particles.

Let us now discuss the high and low field regimes in which there are not
questions related to total spin selection rules.
Figure 2 gives the conductance of the box containing up to 8 electrons for
magnetic fields in two different regimes. For $3$ flux quanta through the box
($B=1.25T$) the many-body wavefunctions are rather complicated including all
the possible spin configurations, while for $12 $ flux quanta ($B=5T$) the
$N$-electrons ground state is spin polarized.
Due to the big differences in the wavefunctions, the peaks behave in a very
different way. For the lower field ({\em i.e.} for non-polarized
wavefunctions) the peak show a monotonous decrease with increasing $N$. On
the contrary, for the higher field ({\em i.e.} for spin polarized
wavefunctions) the spectral weights are such that, after having a
minimum, the peaks tend to have a conductance $e^2/h$.
It must be stressed that we are presenting conductances coming from ground
state
properties at fixed magnetic field, something different to previous
works\cite{514}
where, in order to study the fractional regime,
the system is not maintained in its ground state. Some other minor results
can also be drawn from figure 2: 1) the charging energy is not totally
constant and 2) for the high field, once again, the peak for $2$ electrons is
very
small because it relates $N$-electron wavefunctions with very different
spatial distribution.

In summary, we have computed the conductance of a box in the presence of
a magnetic field at zero temperature including all the electron-electron
interaction without any restriction in the spin configuration.
The conductance consists of a set of peaks with their position and intensity
determined by many-body effects while their width is essentially
given by the coupling between single particle states in the box and in
the leads. In the intermediate range of fields,
the ground state of $N$ electrons change its
multiplet character with the variation of the magnetic field implying
selection rules that quench some conductance peaks. The ground state
wavefunctions
have big differences between the low and the high field regimes. Therefore,
the spectral weights imply very different behavior of the conductance
peaks as a function of the number of electrons in both field regimes.

This work has been supported in part by the Comision Interministerial
de Ciencia y Tecnologia of Spain under contracts MAT 91 0201 and
MAT 91 0905-C02-01 and by the
Commission of the European Communities under contract SSC-CT90-0020.

\newpage

\centerline{FIGURE CAPTIONS}

\vspace{2cm}
Figure 1. Many-body conductance $G$ of a box with $W=100 nm$
and $E_F=11.5meV$ as a function of the dot bottom potential
between $B=2$ and $3 T$.

\vspace{1cm}

Figure 2. Many-body conductance $G$ of a box with $W=100 nm$
and $E_F=11.5meV$ as a function of the dot bottom potential
for $B=1.25$ and $5 T$.
\end{document}